\documentclass[letterpaper,french,english,aps]{revtex4-1}
\usepackage[T1]{fontenc}
\usepackage[latin9]{inputenc}
\usepackage{color}
\usepackage{rotating}
\usepackage{textcomp}
\usepackage{amsmath}
\usepackage{amssymb}
\usepackage{graphicx}

\makeatletter

\providecommand{\tabularnewline}{\\}

\@ifundefined{showcaptionsetup}{}{%
 \PassOptionsToPackage{caption=false}{subfig}}
\usepackage{subfig}
\makeatother

\usepackage{babel}
\makeatletter
\addto\extrasfrench{%
   \providecommand{\fg}{\ifdim\lastskip>\z@\unskip\fi~\frqq}%
}

\makeatother
\begin{document}

\title{Shannon Entropy and Fisher Information of the one-dimensional Klein-Gordon
oscillator with energy-dependent potential}

\author{Abdelmalek Boumali}
\email{boumali.abdelmalek@gmail.com}

\selectlanguage{english}%

\affiliation{Laboratoire de Physique Appliquée et Théorique, ~\\
 Université Larbi-Tébessi- Tébessa, Algeria.}

\author{Malika Labidi}
\email{labidimalika89@gmail.com}

\selectlanguage{english}%

\affiliation{Laboratoire de Physique Appliquée et Théorique, ~\\
 Université Larbi-Tébessi- Tébessa, Algeria.}

\date{\today}
\begin{abstract}
In this paper, we studied, at first, the influence of the energy-dependent
potentials on the one-dimensionless Klein-Gordon oscillator. Then,
the Shannon entropy and Fisher information of this system are investigated.
The position and momentum information entropies for the low-lying
states $n=0,1,2$ are calculated. Some interesting features of both
Fisher and Shannon densities as well as the probability densities
are demonstrated. Finally, the Stam, Cramer\textendash Rao and Bialynicki-Birula\textendash Mycielski
(BBM ) have been checked, and their comparison with the regarding
results have been reported. We showed that the BBM inequality is still
valid in the form $S_{x}+S_{p}\ge1+\text{ln}\pi$ as well as in ordinary
quantum mechanics.
\end{abstract}

\maketitle

\section{Introduction }

Wave equations with energy dependent potentials have been come to
view for long time. They can be seen in Klein-Gordon equation considering
particle in an external electromagnetic field\citep{1}. Arising from
momentum dependent interactions, they also can be appeared in non-relativistic
quantum mechanics, as shown by Green\citep{2} for instance Pauli-Schrödinger
equation possess another example \citep{3,4}. Sazdjian \citep{5}
and Formanek et al. \citep{6} have noted that the density probability,
or the scalar product, has to be modified with respect to the usual
definition, in order to have a conserved norm. Garcia-Martinez et
al \citep{7}. and Lombard \citep{8} made an investigation on Schrödinger
equation with energy-dependent potentials by solving them exactly
in one and three dimensions. Hassanabadi et al. \citep{9} studied
the D-dimensional Schrödinger equation for an energy-dependent Hamiltonian
that linearly depends on energy and quadratic on the relative distance.They
also studied the Dirac equation for an energy-dependent potential
in the presence of spin and pseudospin symmetries with arbitrary spin-orbit
quantum number. They calculate the corresponding eigenfunctions and
eigenvalues of a nonrelativistic energy-dependent system was done
in \citep{10}. A many-body energy-dependent system was studied by
Lombard and Mareš \citep{11}. They considered systems of N bosons
bounded by two-body harmonic interactions, whose frequency depends
on the total energy of the system . Other interesting related works
can be found in \citep{12,13,14,15} and references therein. So, the
Presence of the energy dependent potential in a wave equation has
several non-trivial implications. The most obvious one is the modification
of the scalar product, necessary to ensure the conservation of the
norm. This modification can modified some behavior or physical properties
of a physical system: this question, in best of our knowledge, has
not been considered in the literature.

The relativistic harmonic oscillator is one of the most important
quantum system, as it is one of the very few that can be solved exactly.
The Dirac relativistic oscillator (DO) interaction is an important
potential both for theory and application. It was for the first time
studied by Ito et al\citep{16}. They considered a Dirac equation
in which the momentum $\vec{p}$ is replaced by $\vec{p}-im\beta\omega\vec{r}$,
with $\vec{r}$ being the position vector, $m$ the mass of particle,
and $\omega$ the frequency of the oscillator. The interest in the
problem was revived by Moshinsky and Szczepaniak \citep{17}, who
gave it the name of DO because, in the non-relativistic limit, it
becomes a harmonic oscillator with a very strong spin-orbit coupling
term. Physically, it can be shown that the DO interaction is a physical
system, which can be interpreted as the interaction of the anomalous
magnetic moment with a linear electric field \citep{18,19}. The electromagnetic
potential associated with the DO has been found by Benitez et al\citep{20}.
The DO has attracted a lot of interests both because it provides one
of the examples of the Dirac's equation exact solvability and because
of its numerous physical applications\citep{21,22,23,24,25,26}. \textcolor{black}{Finally,}
Franco-Villafane et al\citep{27} have exposed the proposal of the
first experimental microwave realization of the one-dimensional DO.

The main goal of this paper is studying the effects of the modified
scalar product arising in the energy-dependent Klein-Gordon oscillator
problem. For this, we are focused on the study of: (i) the form of
the spectrum of energy of the one-dimensional Klein-Gordon oscillator
and, (ii) the Fisher and Shannon parameters of quantum information
and the corresponding solutions, and (iii) the validity of Stam \citep{28}
, Cramer\textendash Rao \citep{29,30} and BBM \citep{31} uncertainly
relations for this type of potential.

\section{the one-dimensional Klein-Gordon oscillator with an energy-dependent
potential}

\subsection{The solutions}

\begin{figure}
\subfloat[for particles]{\includegraphics[scale=0.4]{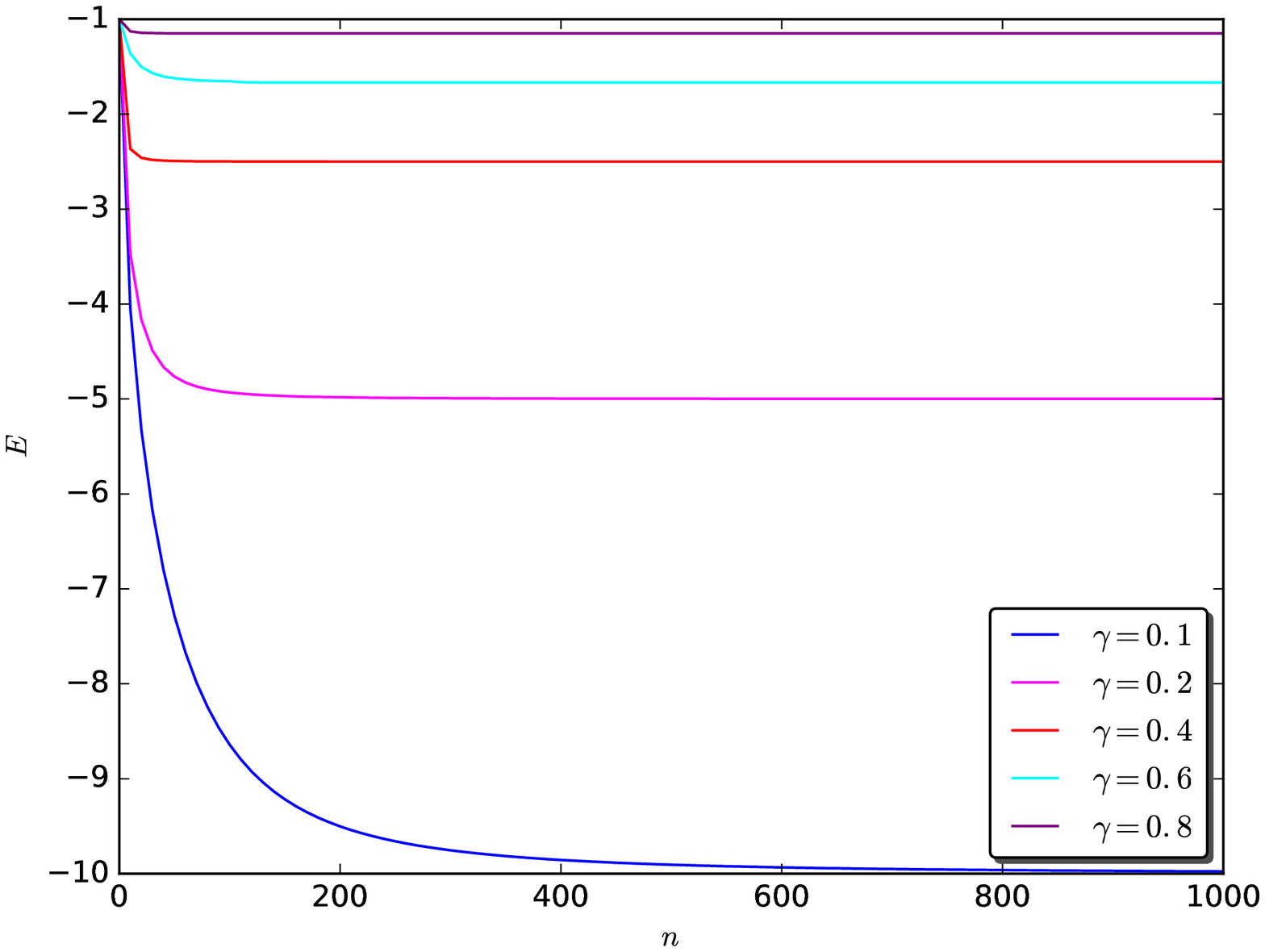}

}\subfloat[for anti-particles]{\includegraphics[scale=0.4]{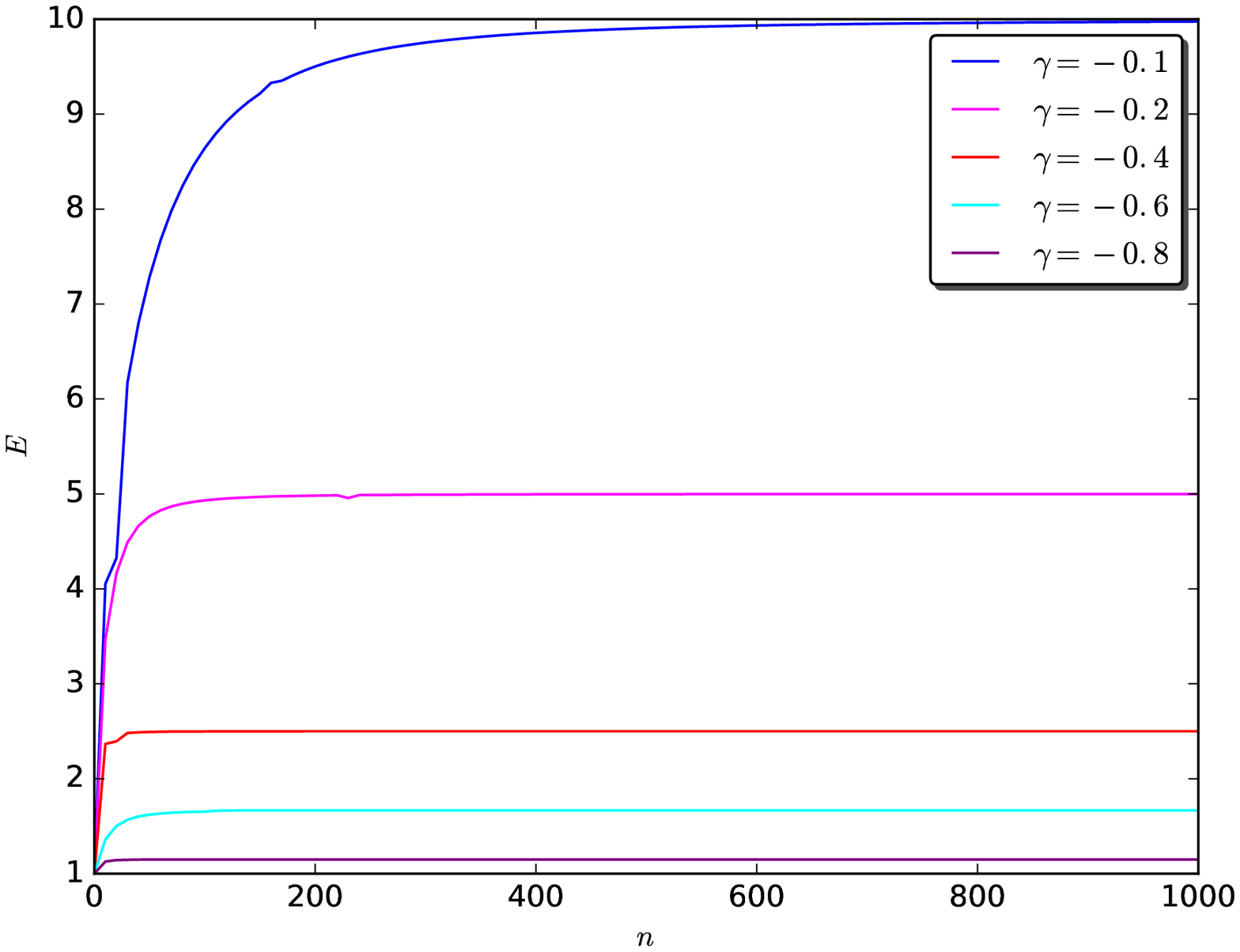}

}

\caption{\label{fig1}Spectrum of energy $E$ versus quantum number $n$ for
some different $\gamma$ values in both coordinate and momentum spaces.}
\end{figure}

In coordinate space, the free Klein-Gordon equation is $\left(\hbar=m=\omega=c=1\right)$
: 
\begin{eqnarray}
\left\{ p_{x}^{2}-\left(E^{2}-1\right)\right\} \psi=0\label{eq:1}
\end{eqnarray}
In the presence of the interaction of the type of Dirac oscillator
$p_{x}\rightarrow p_{x}+ix$ , it becomes \citep{32}:
\[
\left\{ \left(p_{x}+ix\right)\left(p_{x}-ix\right)-\left(E^{2}-1\right)\right\} \psi\left(x\right)=0,
\]
The presence of a potential with energy-dependent potential is shown
by the substituting $p_{x}$ with $p_{x}\rightarrow p_{x}+i\left(1+\gamma E\right)x$
with $\gamma$ is a parameter is not that small and not that big.
In this case, we have 
\begin{equation}
\left(\frac{p_{x}^{2}}{2}+\frac{x^{2}}{2}\right)\psi\left(x,E\right)=\left(\frac{E^{2}-1}{2}+\frac{1}{2}\left(1+\gamma E\right)\right)\psi\left(x,E\right)\label{eq:3}
\end{equation}
With the following substitutions
\begin{equation}
\lambda=\sqrt{1+\gamma E}.\label{eq:4}
\end{equation}
The equation (\ref{eq:3}) represent a equation of a harmonic oscillator
in one-dimensional, and the corresponding eigensolutions are
\begin{equation}
\psi\left(x,E\right)=C_{n}H_{n}\left(\sqrt{\lambda}x\right)\exp\left(-\frac{\lambda}{2}x^{2}\right),\label{eq:5}
\end{equation}
\begin{equation}
E^{4}-2E^{2}-4n^{2}-4n^{2}\gamma E+1=0.\label{eq:6}
\end{equation}
where $C_{n}$ 
\[
\left(C_{n}^{2}\right)=\frac{1}{2^{n}n!}\frac{\left(1+\gamma E\right)^{\frac{1}{4}}}{\sqrt{\pi}}\left(E-\frac{\gamma}{2\sqrt{1+\gamma E}}\left(n+\frac{1}{2}\right)\right)^{-1}.
\]
is the normalization constant, and $H_{n}$ it is the Hermite polynomials.
Now, in momentum space, where we have $p_{x}\rightarrow p_{x}$, and
$x\rightarrow i\frac{\partial}{\partial p_{x}}$, the equation of
Klein-Gordon oscillator has the same form as
\begin{equation}
\left(-\frac{1}{2}\frac{\partial^{2}}{\partial p_{x}^{2}}+\frac{p_{x}^{2}}{2\lambda^{2}}\right)\psi\left(p_{x},E\right)=\frac{E^{2}+\lambda-1}{2\lambda^{2}}\psi\left(p_{x},E\right)\label{eq:7}
\end{equation}
with the corresponding eigensolutions are given by
\begin{equation}
\psi\left(p_{x},E\right)=C'_{n}H_{n}\left(\frac{p_{x}}{\sqrt{\lambda}}\right)\exp\left(-\frac{p_{x}^{2}}{2\lambda}\right),\label{eq:9}
\end{equation}
\begin{equation}
E^{4}-2E^{2}-4n^{2}-4n^{2}\gamma E+1=0.\label{eq:10}
\end{equation}
with 
\[
\left(C_{n}'\right)^{2}=\frac{1}{2^{n}n!\sqrt{\lambda}\sqrt{\pi}}\left\{ E+\frac{\gamma}{2\lambda^{3}}\left(n+\frac{1}{2}\right)\right\} ^{-1}.
\]
The density $\rho_{KG}$ can be expressed by \citep{5}
\begin{align}
\rho_{KG}(x,E) & =\psi\left(x,E\right)^{*}\left(E-\frac{\partial V\left(x,E\right)}{\partial E}\right)\psi\left(x,E\right),\nonumber \\
 & =\left|\psi\left(x,E\right)\right|^{2}\left(E-\frac{1}{2}\gamma x^{2}\right),\label{eq:11}
\end{align}
in the coordinate space. In the momentum space, this form is transformed
into following equation
\begin{align}
\rho_{KG}(p_{x},E) & =\psi\left(p_{x},E\right)^{*}\left(E-\frac{\partial V\left(p_{x},E\right)}{\partial E}\right)\psi\left(p_{x},E\right)\nonumber \\
 & =\left|\psi\left(p_{x},E\right)\right|^{2}\left(E+\frac{\gamma}{2\left(1+\gamma E\right)^{2}}p_{x}^{2}\right).\label{eq:12}
\end{align}
The equations (\ref{eq:6}) and (\ref{eq:10}) are an algebraic equation
of the degree 4 having of the real and complex solutions. The complex
solutions which are not physical, and by the two other real solution
we have plotted Figure. \ref{fig1}. 

In order to represent a physical system, two possibilities, for both
coordinate and momentum spaces, can be made following the sign of
$\rho_{KG}$:
\begin{itemize}
\item if $\rho_{KG}<0$, then we have $\gamma>0$ for the particles $\left(E>0\right)$
\item now, in the other case where $\rho_{KG}>0$, we obtain that have $\gamma>0$
for the anti-particles $\left(E<0\right)$.
\end{itemize}
This imposes constraints on the energy dependence for the theory to
be coherent: by this, we mean a theory that have the following properties:
(i) the necessary modification of the definition of probability density,
(ii) The vectors corresponding to stationary states with different
energies must be orthogonal, (ii) The formulation of the closure rule
in terms of wave functions of stationary states justifies their standardization,
(iv) finally, the operators of observable are all self-adjoint (Hermitian).
In Figure. \ref{fig1}, we have plotted the energy $E$ versus quantum
number $n$ for some different $\gamma$ values in both $\left\{ x\right\} $and
$\left\{ p\right\} $ configuration

Now, we are ready to discuss some interesting results that are not
well comments in the literature. From this Figure, the asymptotic
limits  for both form of energies are $\frac{1}{\left|\gamma\right|}$
as in the non-relativistic case. These limits have been reproduced
for both cases in Figure. \ref{fig1}. Following this figure, some
remarks can be made:
\begin{itemize}
\item the modified scalar product is the origin of that the he spectrum
exhibits saturation instead of growing infinitely, 
\item the analytical asymptotic limits are well depicted, 
\item the beginning of the saturation starts from a specific quantum number
$N$.
\item this saturation appears for the high levels contrary to what has been
found in the non-relativistic case \citep{33}.
\end{itemize}
In what follow, we (i) studied the influence of the dependence of
the potential with energies on the Fisher and Shannon parameters $F$
, and (ii) checked the validity of Stam, Cramer\textendash Rao and
BBM uncertainly relations for some values of $\gamma$.

\section{The influence of the $\gamma$ parameter on the Fisher and Shannon
information measures}

\subsection{Fisher information}

The Fisher information is a quality of an efficient measurement procedure
used for estimating ultimate quantum limits. It was introduced by
Fisher as a measure of intrinsic accuracy in statistical estimation
theory but its basic properties are not completely well known yet,
despite its early origin in 1925. Also, it is the main theoretic tool
of the extreme physical information principle, a general variational
principle which allows one to derive numerous fundamental equations
of physics: Maxwell equations, the Einstein field equations, the Dirac
and Klein-Gordon equations, various laws of statistical physics and
some laws governing nearly incompressible turbulent fluid flows \citep{34,35,36,37,38}.
Fisher information has been very useful and has been applied in different
areas in quantum physics \citep{39,40,41,42,43,44,46,47,48,49,50,51,52}

In our case: the Fisher information of one-dimensional Klein-Gordon
oscillator with energy-dependent potential is
\begin{equation}
F_{x}=\int\rho_{n}\left(x,E\right)\left[\frac{d\text{ln}\left(\rho_{n}\left(x,E\right)\right)}{dx}\right]^{2}dx.\label{eq:13}
\end{equation}
By using the properties of the Hermite functions properties, we found
that
\begin{equation}
\begin{array}{c}
F_{x}=\underset{I}{\underbrace{C_{n}^{2}\int_{-\infty}^{+\infty}\left(16n^{2}E\right)H_{n}^{2}\left(\sqrt{\lambda}x\right)\exp\left(-\lambda x^{2}\right)dx}}+\underset{II}{\underbrace{C_{n}^{2}\int_{-\infty}^{+\infty}-\left(16n\lambda E+8n\gamma\right)x\left(H_{n}^{2}\left(\sqrt{\lambda}x\right)\exp\left(-\lambda x^{2}\right)\right)dx}}+\\
\underset{III}{\underbrace{C_{n}^{2}\int_{-\infty}^{+\infty}\left(4\lambda^{2}E-8n^{2}\gamma+4\gamma\lambda\right)x^{2}\left(H_{n}^{2}\left(\sqrt{\lambda}x\right)\exp\left(-\lambda x^{2}\right)\right)dx}}+\underset{IV}{\underbrace{C_{n}^{2}\int_{-\infty}^{+\infty}8n\lambda\gamma x^{3}\left(H_{n}^{2}\left(\sqrt{\lambda}x\right)\exp\left(-\lambda x^{2}\right)\right)dx}}+\\
\underset{V}{\underbrace{C_{n}^{2}\int_{-\infty}^{+\infty}-2\omega_{E}^{2}\gamma x^{4}\left(H_{n}^{2}\left(\sqrt{\lambda}x\right)\exp\left(-\lambda x^{2}\right)\right)dx}}+\underset{VI}{\underbrace{C_{n}^{2}\int_{-\infty}^{+\infty}\left(\frac{\gamma^{2}x^{2}}{\left(E-\frac{1}{2}\gamma x^{2}\right)}\right)\left(H_{n}^{2}\left(\sqrt{\lambda}x\right)\exp\left(-\lambda x^{2}\right)\right)dx}}
\end{array}\label{eq:15}
\end{equation}
The evaluation of different terms giv
\begin{align*}
I & \rightarrow C_{n}^{2}16n^{2}E2^{n}n!\frac{\sqrt{\pi}}{\sqrt{\lambda}},\\
II & \rightarrow0,\\
III & \rightarrow C_{n}^{2}\left(4\lambda^{2}E-8n^{2}\gamma+4\gamma\lambda\right)\frac{1}{\lambda}2^{n}n!\frac{\sqrt{\pi}}{\sqrt{\lambda}}\left(n+\frac{1}{2}\right),\\
IV & \rightarrow0,\\
V & \rightarrow-C_{n}^{2}2\gamma2^{n}n!\frac{\sqrt{\pi}}{\sqrt{\lambda}}\left[\frac{\left(2n+1\right)^{2}+2}{4}\right].
\end{align*}
The last term
\begin{align*}
VI & \rightarrow C_{n}^{2}\int_{-\infty}^{+\infty}\left(\frac{\gamma^{2}x^{2}}{\left(E-\frac{1}{2}\gamma x^{2}\right)}\right)\left(H_{n}^{2}\left(\sqrt{\lambda}x\right)\exp\left(-\lambda x^{2}\right)\right)dx,
\end{align*}
is calculate numerically.

Hence, the final form of the Fisher parameter is written by :
\begin{equation}
\begin{array}{c}
F_{x}=\left(E-\frac{\gamma}{2\sqrt{1+\gamma E}}\left(n+\frac{1}{2}\right)\right)^{-1}\left\{ 16n^{2}E+\left(4\left(1+\gamma E\right)E-8n^{2}\gamma+4\gamma\sqrt{1+\gamma E}\right)\frac{1}{\sqrt{1+\gamma E}}\left(n+\frac{1}{2}\right)-2\gamma\left[\frac{\left(2n+1\right)^{2}+2}{4}\right]\right\} +\\
\frac{1}{2^{n}n!}\frac{\left(1+\gamma E\right)^{\frac{1}{4}}}{\sqrt{\pi}}\left(E-\frac{\gamma}{2\sqrt{1+\gamma E}}\left(n+\frac{1}{2}\right)\right)^{-1}\int_{-\infty}^{+\infty}\left(\frac{\gamma^{2}x^{2}}{\left(E-\frac{1}{2}\gamma x^{2}\right)}\right)\left(H_{n}^{2}\left(\sqrt{\lambda}x\right)\exp\left(-\lambda x^{2}\right)\right)dx
\end{array}\label{eq:16}
\end{equation}
Let's now go to the momentum space: in this case we have
\begin{equation}
F_{p}=\int\rho_{n}\left(p_{x},E\right)\left[\frac{d\text{ln}\left(\rho_{n}\left(p_{x},E\right)\right)}{dp_{x}}\right]^{2}dp_{x}.\label{eq:17}
\end{equation}
After some calculations, we obtain
\begin{equation}
\begin{array}{c}
F_{p}=\underset{I}{\underbrace{C_{n}^{2}\int_{-\infty}^{+\infty}\left(16n^{2}E\right)H_{n}^{2}\left(\frac{p_{x}}{\sqrt{\lambda}}\right)\exp\left(-\frac{p_{x}^{2}}{\lambda}\right)dp_{x}}}+\underset{II}{\underbrace{C_{n}^{2}\int_{-\infty}^{+\infty}\left(\frac{2\gamma}{\lambda^{4}}-\frac{4nE}{\lambda}\right)p_{x}H_{n}^{2}\left(\frac{p_{x}}{\sqrt{\lambda}}\right)\exp\left(-\frac{p_{x}^{2}}{\lambda}\right)dp_{x}}}+\\
\underset{III}{\underbrace{C_{n}^{2}\int_{-\infty}^{+\infty}\left(\frac{4E}{\lambda^{2}}+\frac{8n^{2}\gamma}{\lambda^{4}}\right)p_{x}^{2}H_{n}^{2}\left(\frac{p_{x}}{\sqrt{\lambda}}\right)\exp\left(-\frac{p_{x}^{2}}{\lambda}\right)dp_{x}}}-\underset{IV}{\underbrace{C_{n}^{2}\int_{-\infty}^{+\infty}\frac{4n\gamma}{2\lambda^{5}}p_{x}^{3}H_{n}^{2}\left(\frac{p_{x}}{\sqrt{\lambda}}\right)\exp\left(-\frac{p_{x}^{2}}{\lambda}\right)dp_{x}}}+\\
\underset{V}{\underbrace{C_{n}^{2}\int_{-\infty}^{+\infty}\frac{4\gamma}{2\lambda^{6}}p_{x}^{4}H_{n}^{2}\left(\frac{p_{x}}{\sqrt{\lambda}}\right)\exp\left(-\frac{p_{x}^{2}}{\lambda}\right)dp_{x}}}+\underset{VI}{\underbrace{C_{n}^{2}\int_{-\infty}^{+\infty}\left(\frac{2\gamma^{2}p_{x}^{2}}{\lambda^{4}\left(2E\lambda^{4}+\gamma p_{x}^{2}\right)}\right)H_{n}^{2}\left(\frac{p_{x}}{\sqrt{\lambda}}\right)\exp\left(-\frac{p_{x}^{2}}{\lambda}\right)dp_{x}}}
\end{array}\label{eq:19}
\end{equation}
When we evaluate the terms appear in Eq. (\ref{eq:19}),
\begin{align*}
I & \rightarrow C_{n}^{2}16n^{2}E2^{n}n!\sqrt{\pi}\sqrt{\lambda},\\
II & \rightarrow0,\\
III & \rightarrow C_{n}^{2}\left(\frac{4E}{\lambda}+\frac{8n^{2}\gamma}{\lambda^{3}}\right)\frac{1}{\lambda}2^{n}n!\sqrt{\pi}\sqrt{\lambda}\left(n+\frac{1}{2}\right),\\
IV & \rightarrow0,\\
V & \rightarrow C_{n}^{2}\frac{2\gamma}{\lambda^{4}}2^{n}n!\sqrt{\pi}\sqrt{\lambda}\left[\frac{\left(2n+1\right)^{2}+2}{4}\right],
\end{align*}
we arrive at the final form of Fisher parameter where
\begin{equation}
\begin{array}{c}
F_{p}=\left(E+\frac{\gamma}{2\lambda^{3}}\left(n+\frac{1}{2}\right)\right)^{-1}\left\{ 16n^{2}E+\left(\frac{4E}{\lambda}+\frac{8n^{2}\gamma}{\lambda^{3}}\right)\left(n+\frac{1}{2}\right)+\frac{2\gamma}{\lambda^{4}}\left[\frac{\left(2n+1\right)^{2}+2}{4}\right]\right\} \\
+\frac{1}{2^{n}n!\sqrt{\lambda}\sqrt{\pi}}\left(E+\frac{\gamma}{2\lambda^{3}}\left(n+\frac{1}{2}\right)\right)^{-1}\int_{-\infty}^{+\infty}\left(\frac{2\gamma^{2}p_{x}^{2}}{\omega_{E}^{4}\left(2E\lambda^{4}+\gamma p_{x}^{2}\right)}\right)H_{n}^{2}\left(\frac{p_{x}}{\sqrt{\lambda}}\right)\exp\left(-\frac{p_{x}^{2}}{\lambda}\right)dp_{x}.
\end{array}\label{eq:20}
\end{equation}
The last term in equation is calculate numerically. 

\subsection{Shannon entropy}

Entropic measures provide analytic tools to help us to understand
correlations in quantum systems. Shannon has introduced entropy to
measure the uncertainty. Now, it has become a universal concept in
statistical physics. The Shannon entropy has finding applications
in several branches of physics because of its possible applications
in a wide range of area (see Ref. \citealp{53} and references therein).

The position space information entropies for the one-dimensional can
be calculated by using 
\begin{equation}
S_{x}=-\int\left|\psi\left(x\right)\right|^{2}\ln\left|\psi\left(x\right)\right|^{2}dx,\label{eq:21}
\end{equation}
In our case, the above equation becomes
\begin{equation}
S_{x}=-\int\rho_{n}\left(x,\gamma\right)\ln\rho_{n}\left(x,\gamma\right)dx,\label{eq:22}
\end{equation}
with $\rho_{n}\left(x,\gamma\right)$ is defined by the equations
(\ref{eq:9}) and (\ref{eq:10}). In general, explicit derivations
of the information entropy are quite difficult. In particular, the
derivation of analytical expression for the $S_{x}$ is almost impossible.The
overcome this difficulties, we (i) use a numerical calculation of
this integral, and (ii) represent the Shannon and Fisher information
entropy densities, respectively,

The form of this parameter, in the momentum space, is written by
\begin{equation}
S_{p}=-\int\rho_{n}\left(p_{x},E\right)\text{ln}\rho_{n}\left(p_{x},E\right)dp_{x}.\label{eq:23}
\end{equation}

\subsection{Results and discussions}

\begin{figure}
\subfloat[coordinate space]{\includegraphics[scale=0.5]{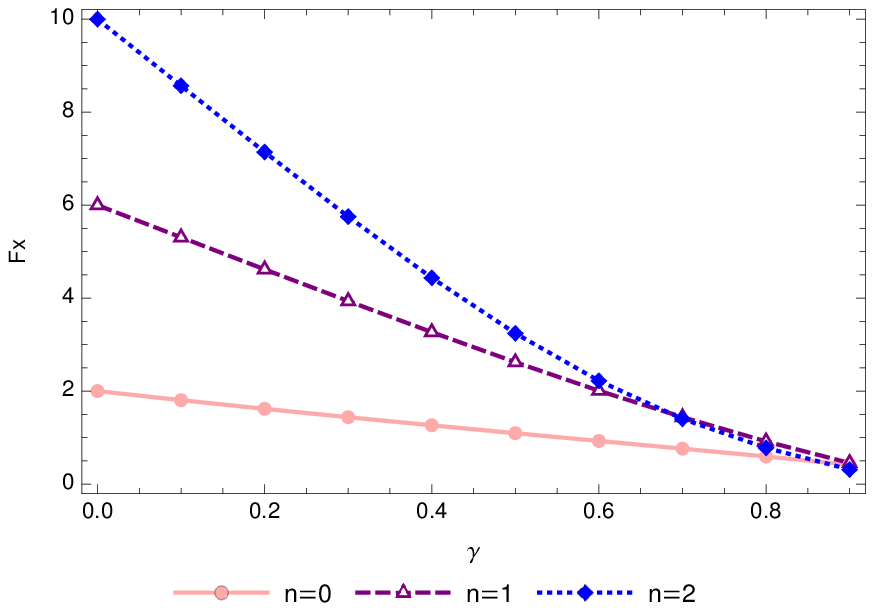}

}\subfloat[momentum space]{\includegraphics[scale=0.5]{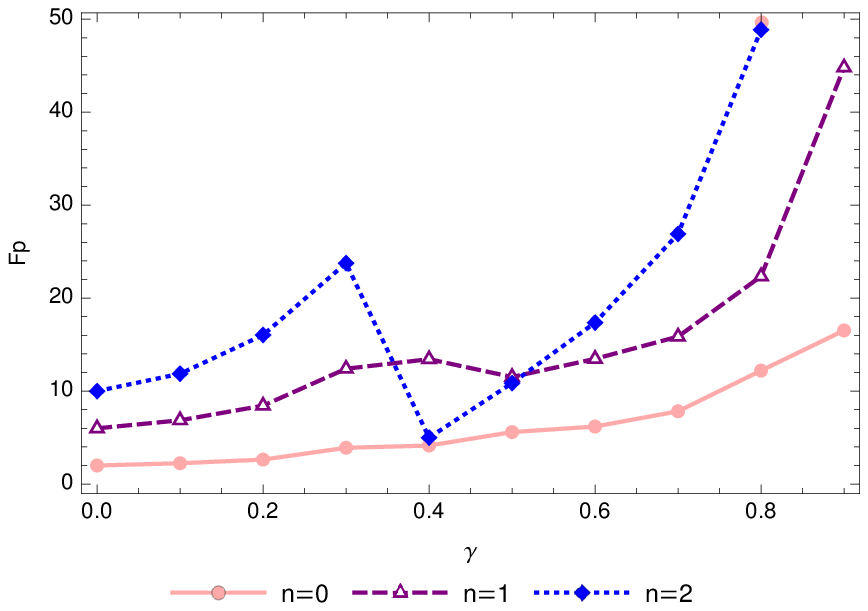}

}

\caption{\label{fig2}Fisher information of one-dimensional Klein-Gordon oscillator
versus $\gamma$ for both coordinate and momentum spaces.}
\end{figure}

\begin{figure}
\subfloat[coordinate space]{\includegraphics[scale=0.5]{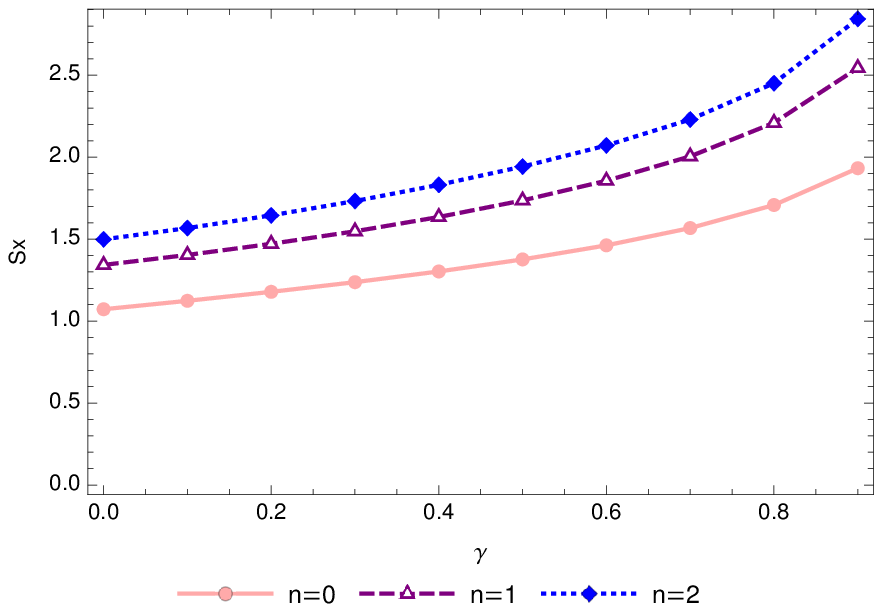}

}\subfloat[momentum ]{\includegraphics[scale=0.5]{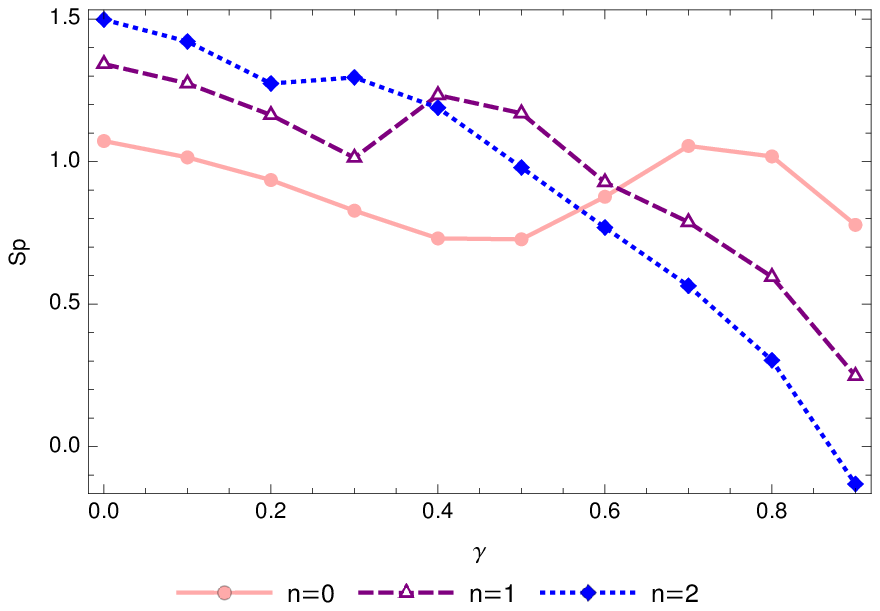}

}

\caption{\label{fig3}Shannon entropy of one-dimensional Klein-Gordon oscillator
versus $\gamma$ for both coordinate and momentum spaces.}
\end{figure}

In Figure. \ref{fig2}, 
 we show the Fisher parameter versus a $\gamma$ for both coordinate
and momentum spaces: the case of coordinate space, $F_{x}$ decreases
contrarily, in the momentum space where it increases. Moreover, this
situation is inverted in the Figure. \ref{fig3} 
: the Shannon parameter increases in the $\left\{ x\right\} $ configuration,
whereas it decreases in the $\left\{ p\right\} $ configuration. We
note here, that these behavior is the same for the particles and anti-particles. 

\begin{figure}
\subfloat[coordinate space]{\includegraphics[scale=0.5]{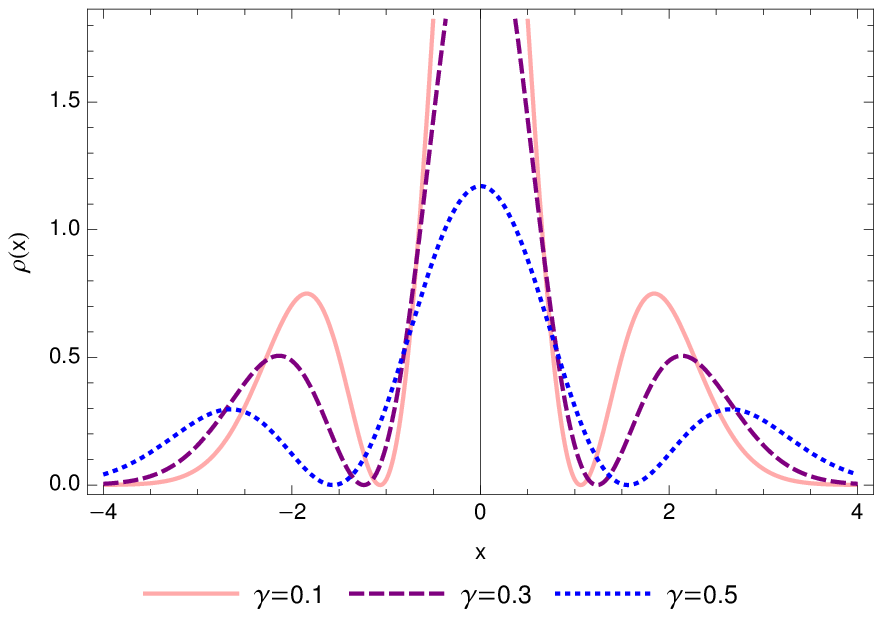}~\includegraphics[scale=0.5]{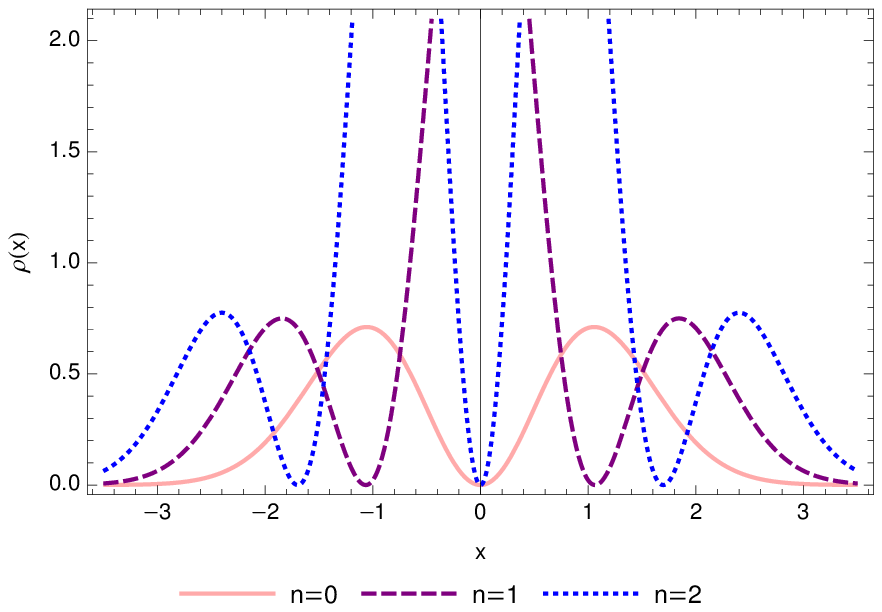}

}

\subfloat[momentum space]{\includegraphics[scale=0.5]{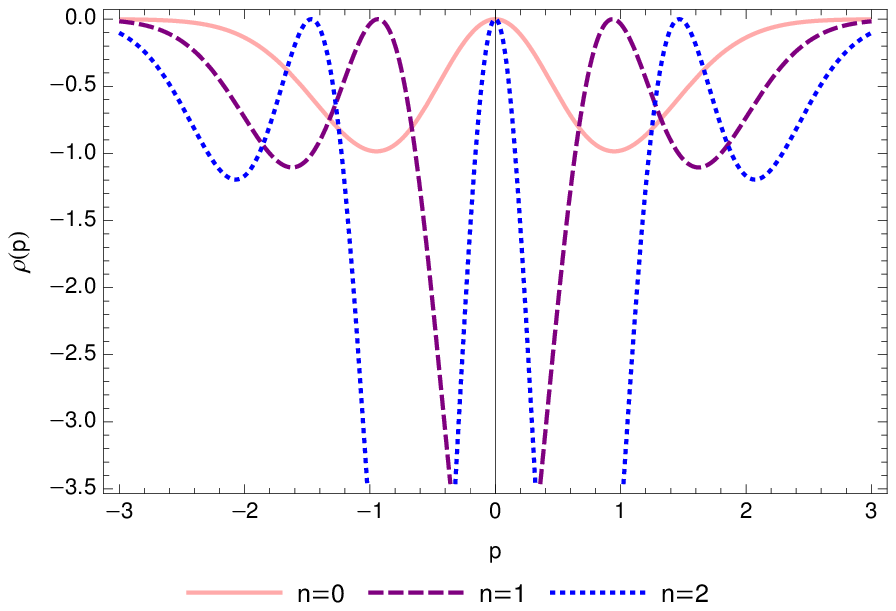} \includegraphics[scale=0.5]{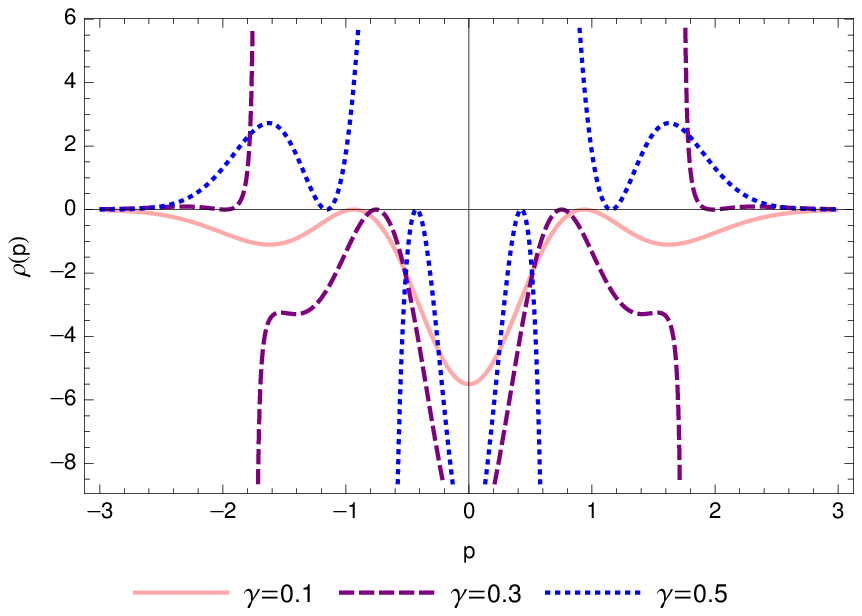}

}

\caption{\label{fig4}Fisher densities versus $x$.}
 
\end{figure}

\begin{figure}
\subfloat[coordinate space]{\includegraphics[scale=0.5]{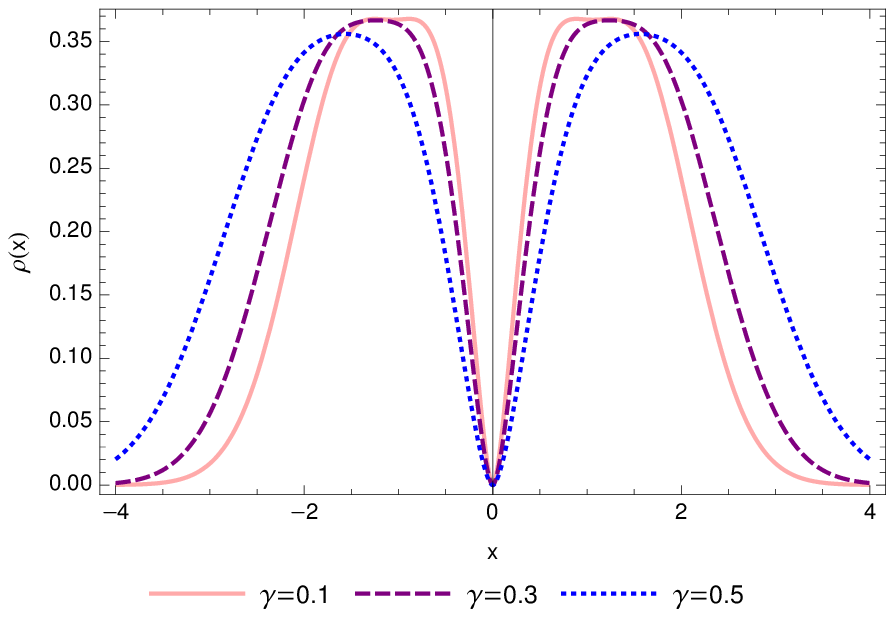} \includegraphics[scale=0.5]{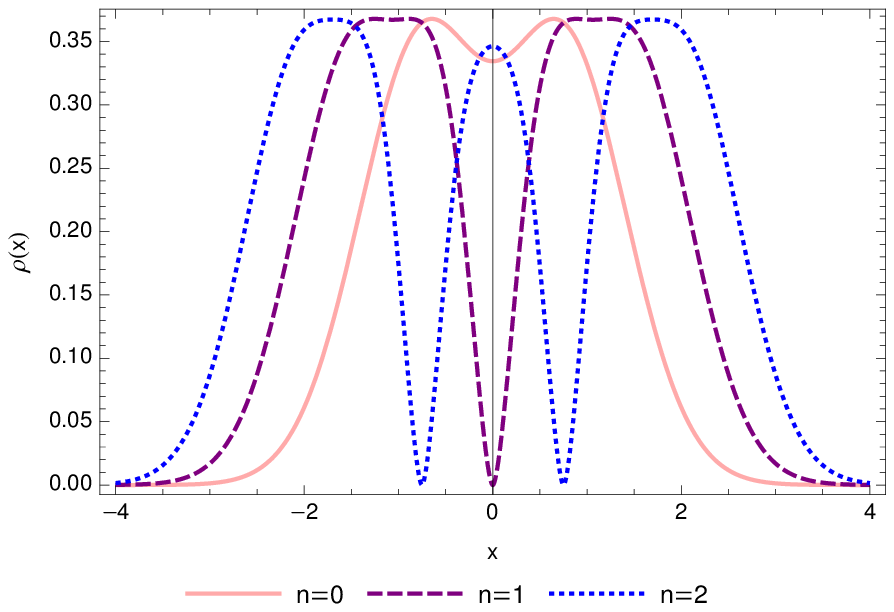}

}

\subfloat[momentum space ]{\includegraphics[scale=0.5]{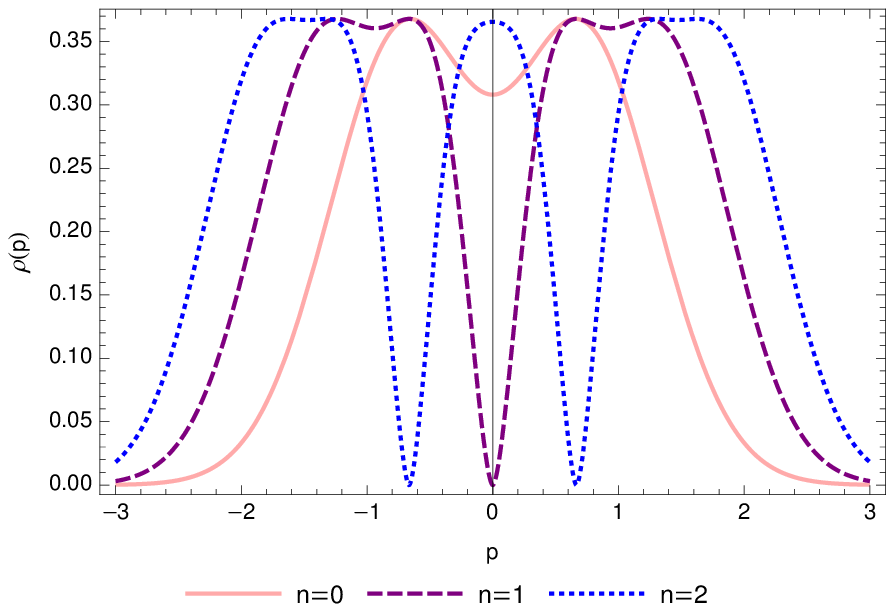} \includegraphics[scale=0.5]{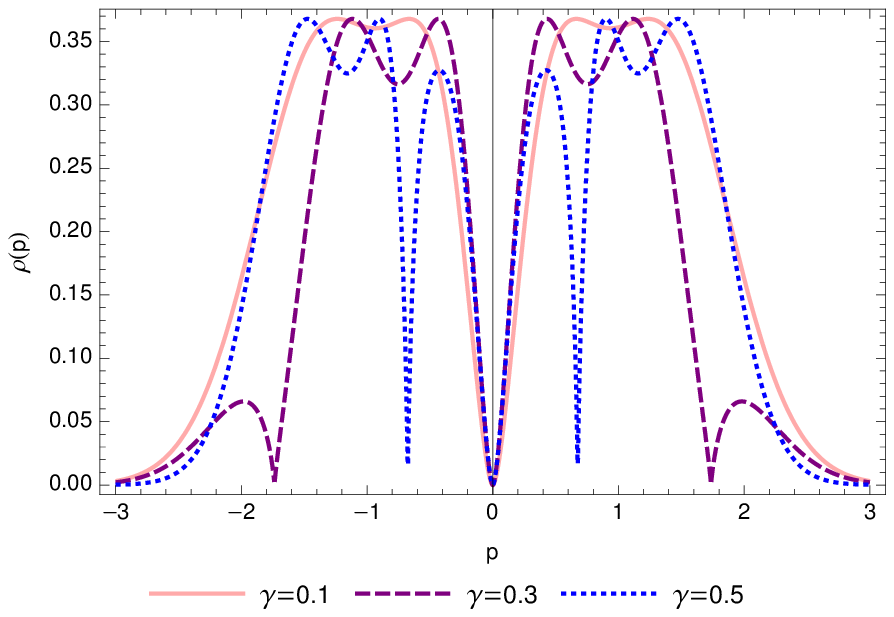}

}

\caption{\label{fig5}Shannon densities versus $p_{x}$.}
\end{figure}

The information entropy and fisher densities are defined as,
\begin{equation}
\left(\rho_{F}\right)_{\left\{ a\right\} }=\rho_{n}\left(a,E\right)\left[\frac{d\text{ln}\left(\rho_{n}\left(a,E\right)\right)}{da}\right]^{2},\label{eq:24}
\end{equation}
for Fisher information, and
\begin{equation}
\left(\rho_{S}\right)_{\left\{ a\right\} }=\rho_{n}\left(a,E\right)\text{ln}\rho_{n}\left(a,E\right),\label{eq:25}
\end{equation}
for the Shannon entropies: $\left\{ a\right\} $ denotes the appropriate
configuration \citealp{54,55}. The behavior of $\left(\rho_{F}\right)_{\left\{ a\right\} }$
and $\left(\rho_{S}\right)_{\left\{ a\right\} }$ is illustrated In
Figures. \ref{fig4} and \ref{fig5} for $n=0,1,3$ and several values
of the parameter $\gamma$ 

Now, we are ready the discuss the Heisenberg uncertainly relation
(HUR) and it's analogue in the framework of quantum information: the
HUR in quantum mechanics is an inequality between position and momentum.
In recent years, a new uncertainty principles are introduced, and
they are originate from the information theory: let us mention that
this information-theoretic quantity and its quantum extension, not
yet sufficiently well known for physicists, has been used to set up
a number of relevant inequalities such as Stam and Cramer\textendash Rao
and uncertainty relations. The Cramer-Rao inequality belongs to a
natural family of information-theoretic inequalities which play a
relevant role in a great variety of scientific and technological fields
ranging from probability theory, communication theory, signal processing
and approximation theory to quantum physics of D -dimensional systems
with a finite number of particles \citep{56,57,58,59,60,61}. 

Also, the Fisher information of single-particle systems has been only
recently determined in closed form in terms of the quantum numbers
characterizing the involved physical state for both position and momentum
spaces. These relevant inequalities which involve the Fisher information
in a given space (Cramer\textendash Rao) or the conjugate (Stam) space
. They are the Stam uncertainty relations \citep{53}
\begin{equation}
F_{x}\leq4\left\langle p^{2}\right\rangle ,\,F_{p}\leq4\left\langle x^{2}\right\rangle ,\label{eq:26}
\end{equation}
and the Cramer\textendash Rao inequalities
\begin{equation}
F_{x}\geq\frac{1}{\left\langle x^{2}\right\rangle },\,F_{p}\geq\frac{1}{\left\langle p^{2}\right\rangle }.\label{eq:27}
\end{equation}
In addition, for a general monodimensional systems we have that 
\begin{equation}
F_{x}F_{p}\geq4.\label{eq:28}
\end{equation}
 Table. \ref{tab:1} 
\begin{table}
\caption{\label{tab:1}Numerical results for the uncertainty relation and Fisher
information and Shannon entropy of 1D Klein-Gordon oscillator with
energy-dependent potential.}

\begin{turn}{90}
\begin{tabular}{|c|c|c|c|c|c|c|c|c|c|c|c|c|}
\hline 
$n$ & $\gamma$ & $\left\langle x^{2}\right\rangle $ & $\triangle x$ & $\left\langle p_{x}^{2}\right\rangle $ & $\triangle p$ & $\triangle x\triangle p$ & $F_{x}$ & $F_{p}$ & $F_{x}F_{p}$ & $S_{x}$ & $S_{p}$ & $S_{x}+S_{p}$\tabularnewline
\hline 
 & 0 & 0.5000 & 0,7071 & 0.5000 & 0,7071 & 0.5000 & 2.0000 & 2.0000 & 4.0000 & 1.0724 & 1.0724 & 2.1448\tabularnewline
\cline{2-13} 
 & -0.16 & 0.5918 & 0,7693  & 0.4080 & 0,638  & 0.4908 & 1.69165  & 2.4516  & 4.1472  & 1.1561  & 0.9707  & 2.1268 \tabularnewline
\cline{2-13} 
 & -0.32 & 0.7136  & 0,8447  & 0.2737  & 0,5232  & 0.4419 & 1.4026  & 3.7700  & 5.2878  & 1.2501  & 0.8054  & 2.0555 \tabularnewline
\cline{2-13} 
0 & -0.48 & 0.8912  & 0,9440  & 0.0188  & 0,1371  & 0.1295 & 1.1276  & 6.9355  & 7.8205  & 1.3608  & 0.7162  & 2.0770 \tabularnewline
\cline{2-13} 
 & -0.64 & 1.1842  & 1,0882  & 0.4410  & 0,6641  & 0.7227 & 0.8607  & 8.2659  & 7.1145  & 1.5014  & 0.9599  & 2.4613 \tabularnewline
\cline{2-13} 
 & -0.80 & 1.8090  & 1,3450  & 0.6658  & 0,8160  & 1.0975 & 0.5927  & 12.2000 & 7.2311  & 1.7082  & 1.0184  & 2.7266 \tabularnewline
\hline 
 & 0 & 1.5000  & 1,2247  & 1,5 & 1,2247   & 1.5000  & 6.0000  & 6.0000  & 36.0000 & 1.3427  & 1.3427  & 2.6854 \tabularnewline
\cline{2-13} 
 & -0.16 & 1.8392  & 1,3562  & 1,17526 & 1,0842  & 1.4704 & 4.8935  & 7.6593  & 37.4804  & 1.4434 & 1.2167  & 2.6601 \tabularnewline
\cline{2-13} 
 & -0.32 & 2.3685  & 1,5390  & 0,480699 & 0,6933  & 1.0670 & 3.8012  & 13.4774  & 51.2303  & 1.5644  & 1.0148  & 2.5792 \tabularnewline
\cline{2-13} 
1 & -0.48 & 3.2837  & 1,8121  & 1,36547 & 1,1685  & 2.1175  & 2.7485  &  6.2457 & 17.1663  & 1.7147  & 1.2308  & 2.9455 \tabularnewline
\cline{2-13} 
 & -0.64 & 5.1536  & 2,2702  & 1,02251 & 1,0112  & 2.2956  & 1.7716  & 13.5291  & 23.9682  & 1.9122  & 0.8693  & 2.7815 \tabularnewline
\cline{2-13} 
 & -0.80 & 10.4112  & 3,2266  & 0,533192 & 0,7302  & 2.3561  & 0.9166  & 22.3000  & 20.4402  & 2.2096  & 0.5957  & 2.8053 \tabularnewline
\hline 
 & 0 & 2,5 & 1,5811  & 2.5000  & 1,5811  & 2.5000  & 10.0000 & 10.0000 & 100.0000 & 1.4986  & 1.4986  & 2.9972 \tabularnewline
\cline{2-13} 
 & -0.16 & 3,23687  & 1,7991  & 1.7994  & 1,3414   & 2.4133  & 7.7103  & 13.8006  & 106.4068  & 1.6128  & 1.3503  & 2.9631 \tabularnewline
\cline{2-13} 
 & -0.32 & 4,51644 & 2,1252  & 0.9625  & 0,9811  & 2.0850  & 5.4827  & 17.7237  & 97.1737  & 1.7513  & 1.3525  & 3.1038 \tabularnewline
\cline{2-13} 
2 & -0.48 & 7,00704 & 2,6471  & 1.7410  & 1,3195  & 3.4928  & 3.4664  & 9.8800  & 34.2480  & 1.9195  & 1.0208  & 2.9403 \tabularnewline
\cline{2-13} 
 & -0.64 & 12,5885 & 3,5480  & 1.0710  & 1,0349  & 3.6718  & 1.8689  & 19.9000  & 37.1911  & 2.1301  & 0.6893  & 2.8194 \tabularnewline
\cline{2-13} 
 & -0.80 & 29,3421 & 5,4168  & 0.4937  & 0,7026  & 3.8085  & 0.7776  & 48.9000  & 38.0261 & 2.2452  & 0.3027  & 2.5479 \tabularnewline
\hline 
\end{tabular}
\end{turn}
\end{table}
shows a numerical results for the uncertainty relation and Fisher
information measure of 1D Klein-Gordon oscillator for three levels
($n=0,1,2$) for some choice of parameter $\gamma$. Following this
Table, we observe that
\begin{itemize}
\item the Stam inequalities, and Cramer\textendash Rao ones are fulfilled, 
\item the following relation 
\begin{equation}
F_{x}F_{p}\geq4,\label{eq:29}
\end{equation}
with D is the space dimension is well-established. 
\item and, finally, as the results indicate that the sum of the entropies
is in consistency with BBM inequality, possesses the stipulated that
\begin{equation}
S_{x}+S_{p}\geq D\left(1+\text{ln}\pi\right).\label{eq:30}
\end{equation}
In our case, we have $S_{x}+S_{p}\geq2.14422$.
\end{itemize}

\section{CONCLUSION}

The present work is devoted to energy dependent potentials. We studied
the physical characteristics of a 1D Klein-Gordon oscillator with
energy dependent-potential. We first obtained the wave functions and
the energy spectra of the system in an exact analytical manner. As
a first result, we showed that the energy dependence affects essentially
the eigenfunctions and eigenvalues. Especially, we observed a saturation
in curves of energy spectrum. Also the presence of the energy\textendash dependent
potential in a wave equation leads to the modification of the scalar
product, which was necessary to ensure the conservation of the norm.
In this context, Fisher information and Shannon entropy, some expectation
values, and some uncertainty principles were evaluated: in this way,
we have studied the influence of the parameter $\gamma$ on Shannon
entropy and Fisher information uncertainty relations, and checked
the validity of BBM inequality. We showed that the numerical results
in the information entropic is predicted by the BBM inequality $S_{x}+S_{p}\geq1+\text{ln}\pi$,
for some values of parameter $\gamma$ . In conclusion, the uncertainly
relations given by quantum information theory, can be extended normally
to the case of the potentials which depend with energy.


\begin{thebibliography}{99}
\bibitem[1]{1}H. Snyder and J. Weinberg, Phys. Rev. \textbf{57},
307 (1940);\\
 I. Schiff, H. Snyder and J.Weinberg, Phys. Rev. \textbf{57}, 315
(1940).

\bibitem[2]{2}A.M. Green, Nucl. Phys. \textbf{33} ,218 (1962).

\bibitem[3]{3}W. Pauli, Z. Physik. \textbf{601}, 43 (1927).

\bibitem[4]{4}H.A. Bethe and E.E. Salpeter, Quantum theory of One-
and Two-Electron Systems, Handbuch der Physik, Band XXXV, Atome I,
Springer Verlag, Berlin-G\textasciidieresis ottingen-Heidelberg, (1957).

\bibitem[5]{5}H. Sazdjian, J. Math. Phys. \textbf{29}, 1620 (1988).

\bibitem[6]{6}J. Formanek, J. Mares and R. Lombard, Czech. J. Phys.
\textbf{54}, 289 (2004).

\bibitem[7]{7}J. Garcia-Martinez, J. Garcia-Ravelo, J. J. Pena and
A. Schulze-Halberg, Phys. Lett. A. \textbf{373}, 3619 (2009).

\bibitem[8]{8}R. Lombard, An-Najah Univ. J. Res. (N. Sc.). \textbf{25},
49 (2011).

\bibitem[9]{9}H. Hassanabadi, S. Zarrinkamar and A. A. Rajabi, Commun.
Theor. Phys. \textbf{55}, 541 (2011).

\bibitem[10]{10}H. Hassanabadi, E. Maghsoodi, R. Oudi, S. Zarrinkamar,
H. Rahimov, Eur. Phys. J. Plus. \textbf{127}, 120 (2012).

\bibitem[11]{11}R. J. Lombard and J. Mares, Phys. Lett. A. \textbf{373},
426 (2009).

\bibitem[12]{12}R. Yekken and R. J. Lombard, J. Phys. A: Math. Theor.
\textbf{43}, 125301 (2010);\\
 R. Yekken, Phd Thesis, \textbf{\emph{Université des Sciences et de
la Technologie Houari Boumediene d'Alger}}, Algeria, (2009).

\bibitem[13]{13}A. Schulze-Halberg, Cent. Eur. J. Phys. \textbf{9},
57 (2011).

\bibitem[14]{14}R. J. Lombard, J. Mares and C. Volpe, arXiv:hep-ph/0411067v1.

\bibitem[15]{15}H. Hassanabadi, S. Zarrinkamar, H. Hamzavi and A.
A. Rajabi, Arab. J. Sci. Eng. \textbf{37}, 209 (2012).

\bibitem[16]{16}\foreignlanguage{french}{D. Itô, K. Mori and E. Carriere,
Nuovo Cimento A, \textbf{51}, 1119 (1967). }

\selectlanguage{french}%
\bibitem[17]{17}M. Moshinsky and A. Szczepaniak, J. Phys. A: Math.
Gen, \textbf{22,} L817 (1989).

\bibitem[18]{18}R. P. Martinez-y-Romero and A. L. Salas-Brito, J.
Math. Phys, \textbf{33} , 1831 (1992).

\bibitem[19]{19}M. Moreno and A. Zentella, J. Phys. A: Math. Gen,
\textbf{22} , L821 (1989).

\bibitem[20]{20}J. Benitez, P. R. Martinez y Romero , H. N. Nunez-Yepez
and A. L. Salas-Brito,Phys. Rev. Lett, \textbf{64}, 1643\textendash 5
(1990).

\bibitem[21]{21}C. Quesne and V. M. Tkachuk, J. Phys. A: Math. Gen,
\textbf{41} , 1747\textendash 65 (2005).

\selectlanguage{english}%
\bibitem[22]{22}\foreignlanguage{french}{A. Boumali and H. Hassanabadi,
Eur. Phys. J. Plus. \textbf{128}, 124 (2013).}

\bibitem[23]{23}\foreignlanguage{french}{A. Boumali and H. Hassanabadi,
Z. Naturforschung A. \textbf{70}, 619-627 (2015).}

\bibitem[24]{24}A. Boumali, EJTP \textbf{12}, No. 32,1-10 (2015).

\bibitem[25]{25}A. Boumali, Phys. Scr. \textbf{90}, 045702 (2015).

\bibitem[26]{26}\textcolor{black}{P. Strange, L.H. Ryder, Phys. Lett.
A , }\textbf{\textcolor{black}{380}}\textcolor{black}{, 3465-3468
(2016).}

\bibitem[27]{27}\foreignlanguage{french}{A. Franco-Villafane, E.
Sadurni, S. Barkhofen, U. Kuhl, F. Mortessagne, and T. H. Selig- man,
Phys. Rev. Lett. \textbf{111}, 170405 (2013).}

\bibitem[28]{28}A. J. Stam A J Inf. Control \textbf{2}, 101 (1959).

\bibitem[29]{29}T. M. Cover and J. A.Thomas, Elements of Information
Theory,Wiley, NewYork,(1991). 

\bibitem[30]{30}O. Johnson, Information Theory and the Central Limit
Theorem, Imperial College Press, London (2004).

\bibitem[31]{31}Iwo Biatynicki-Birula and Jerzy Mycielski, Commun,
math. Phys. \textbf{44}, 129-{}-132 (1975). 

\bibitem[32]{32}A. Boumali, A. Hafdallah and A. Toumi, Phys. Scr.
\textbf{84}, 037001 (2011).

\bibitem[33]{33}A. Boumali, S. Dilmi, S. Zare and H. Hassanabdi,
arXiv:1607.02123v2 (2016)

\bibitem[34]{34}B.R. Frieden, Am. J. Phys. \textbf{57}, 1004 (1989).

\bibitem[35]{35}B.R. Frieden, Phys. Rev. A. \textbf{41}, 4265 (1990).

\bibitem[36]{36}B.R. Frieden, Opt. Lett. \textbf{14}, 199 (1989).

\bibitem[37]{37}B. R. Frieden, Physica. A. \textbf{180}, 359-385
(1992).

\bibitem[38]{38}B. R. Frieden and B. H. Soffer, Phys. Rev. E. \textbf{52},
2274-2286 (1995).

\bibitem[39]{39}D.X. Macedo and I. Guedes, Physica. A. \textbf{434},
211-219 (2015).(2008).

\bibitem[40]{40}P.A. Bouvrie, J.C. Angulo and J.S. Dehesa, Physica.
A. \textbf{390}, 2215-2228 (2011).

\bibitem[41]{41}J. S. Dehesa, S. López-Rosa, and B. Olmos, J. Math.
Phys. \textbf{47}, 052104 (2006).

\bibitem[42]{42}B J Falaye, K J Oyewumi, S M Ikhdair and M Hamzavi,
Phys. Scr. \textbf{89} , 115204 (2014).

\bibitem[43]{43}V. Aguiar and I. Guedes, Phys. Scr. \textbf{90},
045207 (2015).

\bibitem[44]{44}B.J. Falayea, F.A. Serranob and Shi-Hai Dongc, Phys.
Lett. A. \textbf{380}, 267\textendash 271 (2016).

\bibitem[45]{45}Juan He , Zhi-Yong Ding and Liu Ye, Physica. A, (2016).

\bibitem[46]{46}M. Ghafourian and H. Hassanabadi, J. Korean. Phys.
Soc, \textbf{68}, 1-4 (2016),

\bibitem[47]{47}G. H. Sun, S. H. Dong and S. Naad, Ann. Phys. \textbf{525},
934 (2013).

\bibitem[48]{48}G. H. Sun, M. Avila Aoki and S. H. Dong, Chin. Phys.
B \textbf{22}, 050302 (2013).

\bibitem[49]{49}S. Dong, G. H. Sun, S. H. Dong and J. P. Draayer,
Phys. Lett. A \textbf{378}, 124 (2014).

\bibitem[50]{50}G. H. Sun, S.H. Dong, K. D. Launey, T. Dytrych and
J. P., Draayer, Int. J. Quantum Chem. \textbf{115}, 891 (2015).

\bibitem[51]{51}R. Valencia-Torres, G. H. Sun and S. H. Dong, Phys.
Scr. \textbf{90}, 035205 (2015).

\bibitem[52]{52}B. J. Falaye, F. A. Serrano and S. H. Dong, Phys.
Lett. A \textbf{380}, 267 (2016). 

\bibitem[53]{53}J S Dehesa, R González-Férez and P Sánchez-Moreno,
J. Phys. A: Math. Theor. \textbf{40},1845 (2007).

\bibitem[54]{54}J-F Bercher, J. Phys. A: Math. Theor. \textbf{45},
255303 (2012).

\bibitem[55]{55}S. Zare and H. Hassanabadi, Advances in High Energy
Physics Volume 2016, Article ID 4717012,

\bibitem[56]{56}Aparna Saha, B Talukdar and Supriya Chatterjee, Eur.
J. Phys. \textbf{38}, 025103 (2017).

\bibitem[57]{57}D Manzano 1,2,4 , R J Yáñez 1,3,4 and J S Dehesa,
New. J. Phys. \textbf{12}, 023014 (2010).

\bibitem[58]{58}Jaime Sañudo and Ricardo López-Ruiz, J. Phys. A:
Math. Theor. \textbf{41}, 265303 (2008). 

\bibitem[59]{59}P Sánchez-Moreno, R González-Férez and J S Dehesa,
New. J. Phys. \textbf{8}, 330 (2006).(

\bibitem[60]{60}Seyede Amene Najafizade, Hassan Hassanabadi, and
Saber Zarrinkamar, Can. J. Phys\textbf{. 94}, 1085-1092 (2016). 

\bibitem[61]{61}J.S. Dehesa, A.R. Plastino, P. Sanchez-Moreno, C.
Vignat, Appl. Math. Lett. \textbf{25}, 1689\textendash 1694 (2012).
\end{thebibliography}
\end{document}